\documentstyle[prl,aps,twocolumn,psfig,floats]{revtex}

\begin{document}

\title{Nuclear multifragmentation, percolation and the Fisher Droplet Model: common features of reducibility and thermal scaling}

\author{
J. B. Elliott$^{ 1 }$, 
L. G. Moretto$^{ 1 }$, 
L. Phair$^{ 1 }$, 
G. J. Wozniak$^{ 1 }$,\\
S. Albergo$^{ 2 }$, 
F. Bieser$^{ 1 }$, 
F. P. Brady$^{ 3 }$, 
Z. Caccia$^{ 2 }$, 
D. A. Cebra$^{ 3 }$, 
A. D. Chacon$^{ 4 }$, 
J. L. Chance$^{ 3 }$,\\
Y. Choi$^{ 5 }$, 
S. Costa$^{ 2 }$, 
M. L. Gilkes$^{ 5 }$, 
J. A. Hauger$^{ 5 }$, 
A. S. Hirsch$^{ 5 }$, 
E. L. Hjort$^{ 5 }$, 
A. Insolia$^{ 2 }$, 
M. Justice$^{ 6 }$,\\
D. Keane$^{ 6 }$, 
J. C. Kintner$^{ 3 }$, 
V. Lindenstruth$^{ 7 }$, 
M. A. Lisa$^{ 1 }$, 
H. S. Matis$^{ 1 }$, 
M. McMahan$^{ 1 }$, 
C. McParland$^{ 1 }$,\\
W. F. J. M\"{u}ller$^{ 7 }$,
D. L. Olson$^{ 1 }$, 
M. D. Partlan$^{ 3 }$, 
N. T. Porile$^{ 5 }$, 
R. Potenza$^{ 2 }$, 
G. Rai$^{ 1 }$, 
J. Rasmussen$^{ 1 }$,\\
H. G. Ritter$^{ 1 }$,
J. Romanski$^{ 2 }$, 
J. L. Romero$^{ 3 }$, 
G. V. Russo$^{ 2 }$, 
H. Sann$^{ 7 }$, 
R. P. Scharenberg$^{ 5 }$, 
A. Scott$^{ 6 }$, 
Y. Shao$^{ 6 }$,\\													
B. K. Srivastava$^{ 5 }$, 
T. J. M. Symons$^{ 1 }$, 
M. Tincknell$^{ 5 }$, 
C. Tuv\'{e}$^{ 2 }$, 
S. Wang$^{ 6 }$, 
P. G. Warren$^{ 5 }$, 
H. H. Wieman$^{ 1 }$,\\
T. Wienold$^{ 1 }$, and 
K. Wolf$^{ 4 }$\\
}

\address{
$^1$Nuclear Science Division, Lawrence Berkeley National Laboratory, Berkeley, CA 94720\\
$^2$Universit\'{a} di Catania and Istituto Nazionale di Fisica Nucleare-Sezione di Catania, 95129 Catania, Italy\\ 
$^3$University of California, Davis, CA 95616\\ 
$^4$Texas A\&M University, College Station, TX  77843\\
$^5$Purdue University, West Lafayette, IN 47907\\
$^6$Kent State University, Kent, OH 44242\\
$^7$GSI, D-64220 Darmstadt, Germany
}

\date{\today}

\maketitle

\begin{abstract}
It is shown that the Fisher Droplet Model (FDM), percolation and nuclear multifragmentation share the common features of 
reducibility (stochasticity in multiplicity distributions) and thermal scaling (one-fragment production probabilities are 
Boltzmann factors).\ \ Barriers obtained, for cluster production on percolation lattices, from the Boltzmann factors show a 
power-law dependence on cluster size with an exponent of $0.42 \pm 0.02$.\ \ The EOS Au multifragmentation data yield barriers 
with a power-law exponent of $0.68 \pm 0.03$.\ \ Values of the surface energy coefficient of a low density nuclear system are 
also extracted.\\
\end{abstract}

\pacs{25.70 Pq, 64.60.Ak, 24.60.Ky, 05.70.Jk}

\narrowtext

Since the earliest observations of nuclear multifragmentation (the break up of excited nuclei), the Fisher Droplet Model (FDM) 
\cite{fisher} and percolation models \cite{stauffer_aharony} have been employed in attempts to understand this phenomenon.\ \ 
The FDM enjoyed early success in predicting power-law distributions in fragment masses at the critical point in a liquid-vapor 
diagram \cite{finn}.\ \ Percolation models also predicted a power-law distribution in fragment sizes near the critical point 
\cite{campi_2}, \cite{bauer_1}.\ \ Both models still enjoy great popularity and have been employed in the analysis of Au 
multifragmentation data obtained by the EOS Collaboration 
\cite{gilkes_gamma},\cite{elliott_sigma},\cite{hauger_prl},\cite{hauger_prc},\cite{elliott_scaling}.

Other analyses of multifragmentation data have shown two empirical properties of the fragment multiplicities which have been 
named reducibility and thermal scaling \cite{moretto_rev},\cite{beaulieu},\cite{toke},\cite{phair}.\ \ Reducibility refers to 
the observation that for each energy bin, $E$, the fragment multiplicities, $N$, are distributed according to a binomial or 
Poissonian law.\ \ As such, their multiplicity distributions, $P_N$, can be {\it reduced} to a {\it one-fragment production 
probability} $p$, according to the binomial or Poissonian law:
	\begin{eqnarray} 
	P^{M}_{N} & = & \frac{M!}{M!(M-N)!} p^{N} (1-p)^{M-N} ; \nonumber \\
	P_{N} & = & e^{-{\left< N \right>}} \frac{1}{N!} {\left< N \right>}^{N} ,
	\label{bino} 
	\end{eqnarray}
where $M$ is the total number of trials in the binomial distribution.\ \ The experimental observation that $P_N$ could be 
constructed in terms of $p$ was considered evidence for stochastic fragment production, {\it i.e.} fragments are produced 
independently of each other.\ \ Experimental fragment multiplicity distributions were observed to change from binomial to 
Poissonian under a redefinition of fragment from $3 \le Z \le 20$ to individual charges, $Z$ \cite{beaulieu}.

Thermal scaling refers to the feature that $p$ behaves with temperature $T$ as a Boltzmann factor: $p \propto \exp( -B / T )$.\ \ 
Thus a plot of $\ln p$ vs. $1 / T$, an Arrhenius plot, should be linear if $p$ is a Boltzmann factor.\ \ The slope $B$ is the 
one-fragment production barrier.\ \ Analyses of multifragmentation distributions along these lines have demonstrated the presence 
of these features \cite{moretto_rev} and have led to the extraction of barriers \cite{beaulieu}.\ \ Controversy has surrounded this 
type of analysis regarding both the physical existence of these features and their significance, mostly within the framework of 
dynamical vs. statistical origins of multifragmentation \cite{toke}.

In this work several important points will be made: The FDM inherently contains reducibility and thermal scaling.\ \ Since 
percolation reduces to the FDM, it exhibits reducibility and thermal scaling.\ \ Thus percolation provides a simple mathematical 
model that fully manifests these two features.\ \ Arrhenius plots for percolation can be used to extract barriers.\ \ The barriers 
have a power-law dependence on cluster size.\ \ Analysis of the EOS Au multifragmentation data verifies reducibility and thermal 
scaling.\ \ The extracted barriers also obey a power-law dependence on fragment mass.

The FDM and its forerunners \cite{frenkel}, \cite{mayer} are based on the equilibrium description of physical clusters or 
droplets.\ \ The mean number of droplets of size $A$ was written as:
	\begin{equation}
	\left< N_A \right> \propto \exp \left[ \frac{A {\Delta}{\mu} }{ T }\right] , 
	\label{cluster1}
	\end{equation}
where ${\Delta}{\mu} = \mu - {\mu}_l$ and $\mu$ and ${\mu}_l$ are the actual and liquid chemical potentials respectively.\ \ For 
$\mu < {\mu}_l$ (gas), $\left< N_A \right>$ falls to zero with increasing $A$.\ \ For $\mu > {\mu}_l$ (liquid), 
$\left< N_A \right>$ increases with $A$.\ \ To better describe the distribution for intermediate values of $A$, 
Eq.~(\ref{cluster1}) was modified to include the surface of the droplets:
	\begin{equation}
	\left< N_A \right> \propto \exp \left[ \frac{A{\Delta}{\mu}}{T} - \frac{ c(T) A^{2/3} }{ T} \right] ,
	\label{cluster2}
	\end{equation}
where $c(T)$ is the surface free-energy density.\ \ For $\mu < {\mu}_l$, $\left< N_A \right>$ falls to zero 
with increasing $A$.\ \ For $\mu > {\mu}_l$, the terms in the exponential compete, leading to an early decrease in 
$\left< N_A \right>$ with $A$, followed by an increase.

To account for the properties near criticality, Fisher introduced an explicit expression for the surface free energy and a 
topological factor resulting in an expression for the {\it normalized} droplet distribution:
	\begin{equation}
	\left< n_A \right> = \left< \frac{N_A}{A_0} \right> = 
			     q_0 A^{-\tau} \exp \left[ \frac{A{\Delta}{\mu} }{ T } - 
                                                       \frac{c_0 \epsilon A^{\sigma} }{ T } \right] ,
	\label{fisher_droplet}
	\end{equation}
where: $A_0$ is the size of the system; $q_0$ is a normalization constant depending only on the value of $\tau$ \cite{nakanishi}; 
$\tau$, the topological critical exponent, depends on the dimensionality of the system with origins that lie in considerations of 
a three dimensional random walk of a surface closing on itself, for three dimensions $2 \le {\tau} \le 3$; $c_0 \epsilon A^{\sigma}$ 
is the surface free energy of a droplet of size $A$; $c_0$ is the surface energy coefficient; $\sigma$ is the critical exponent 
related to the ratio of the dimensionality of the surface to that of the volume; and $\epsilon = ( T_c - T ) / T_c$ is the control 
parameter, a measure of the distance from the critical point, $T_c$.

From this outline it is apparent that the FDM exhibits the features of reducibility and thermal scaling.\ \ The distribution in 
droplet size is Poissonian by construction: in the FDM each component of droplet size $A$ is an ideal gas without the canonical 
constraint of overall constituent number conservation.\ \ The resulting grand canonical distribution is Poissonian.\ \ Thus, 
${\sigma}^{2}_{A} = \left< N_A \right>$, {\it i.e.} Poissonian reducibility.

Thermal scaling is obvious in the FDM when Eq.~(\ref{fisher_droplet}) is written as follows:
	\begin{equation}
	\ln \left< n_A \right> = \ln q_0 - \tau \ln A 
                               + \frac{A{\Delta}{\mu}}{T}
                               + \frac{c_0 A^{\sigma}}{T_c} 
                               - \frac{c_0 A^{\sigma}}{T} .
	\label{arrhenius_fisher}
	\end{equation}
It is clear that linearity with $1/T$ (thermal scaling in an Arrhenius plot) extends to and beyond the critical point, and the slope 
of the Arrhenius plot gives the $T=0$ surface energy coefficient of the droplet.

Percolation models are characterized by a constant energy per bond.\ \ The bond-breaking probability, $p_{break}$, is amenable to a 
straightforward statistical mechanics treatment.\ \ Such a treatment reveals that in the limit of $T \rightarrow \infty$, 
$p_{break} \rightarrow 1/2$, indicating that the range of $p_{break}$ covered by Eq.~(\ref{fisher_droplet}) is half the usual range 
discussed in percolation theory; $0 \le p_{break} \le 1$.\ \ Such a literal thermodynamical treatment therefore excludes the critical 
point, $p_c$, of many types of percolation systems from thermodynamic consideration.

However, percolation phenomena, with a geometrical phase transition, share with thermal critical phenomena the important features of 
{\it scaling}, {\it universality} and {\it renormalization group} as well as other deep connections \cite{stanley}.\ \ For example, 
the scaling behavior observed in percolation clusters can be described by the FDM when $p_{break}$ replaces $T$ in 
Eq.~(\ref{fisher_droplet}) and the control parameter becomes $\epsilon = ( p_c - p_{break} ) / p_c$ \cite{stauffer_aharony}.

\begin{figure} [ht]
\centerline{\psfig{file=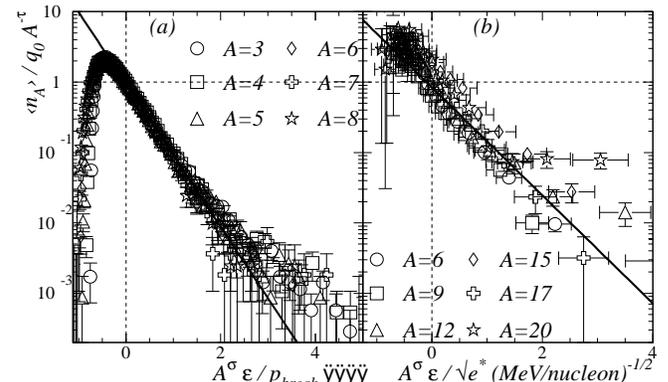,width=8.5cm,angle=0}}
\caption{The scaled cluster (fragment) distribution plotted as a function of the scaled control parameter for clusters (fragments) 
of size (mass) $A$.\ \ The solid line shows a fit to the FDM for (a) percolation ((b) Au multifragmentation).}
\label{perc_scal}
\end{figure}

To demonstrate the scaling of percolation clusters a plot is made of the scaled cluster distribution, 
$n^{scaled}_{A} = \left< n_A \right> / q_0 A^{-\tau}$, as a function of the scaled control parameter, 
${\epsilon}^{scaled} = A^{\sigma} \epsilon / p_{break}$.\ \ See the left panel of Fig.~\ref{perc_scal}.\ \ Data over a wide range 
in $A$ and $\epsilon$ are seen to collapse.\ \ Cluster distributions used in this analysis were generated on a simple cubic lattice 
of side six \cite{elliott_perc_2}.

Fitting $n^{scaled}_{A}$ as a function of ${\epsilon}^{scaled}$ for $\epsilon \ge 0$ and leaving $c_0$ and 
$\exp \left[ A {\Delta}{\mu} / p_{break} \right]$ as free parameters in Eq.~(\ref{fisher_droplet}) gives $c_0 = 2.34 \pm 0.03$ 
and $\exp \left[ A{\Delta}{\mu} /p_{break} \right] = 0.95 \pm 0.01$, {\it i.e.} the bulk factor is unity.\ \ At the critical point, 
$\epsilon = 0$, the collapsed distribution takes the value of one indicating that the cluster distribution follows a power law.\ \ 
Away from the critical point, the cluster distribution predominantly follows the surface term in Eq.~(\ref{fisher_droplet}).\ \ The 
FDM does not describe the behavior of clusters for $\epsilon \ll 0$.\ \ Other forms for the FDM's surface factor have been suggested 
to describe cluster behavior on both sides of the critical point \cite{stauffer_aharony}, \cite{elliott_perc_2}.

Also shown in Fig.~\ref{perc_scal} is a plot of the EOS Au multifragmentation data \cite{gilkes_gamma}, \cite{elliott_sigma}, 
\cite{hauger_prl}, \cite{hauger_prc}, \cite{elliott_scaling}.\ \ Here the substitution of $\sqrt{e^*} = \sqrt{E^{*}/A_0}$ for $T$ 
has been made resulting in a control parameter of $\epsilon = (\sqrt{e^{*}_{c}} - \sqrt{e^*} ) / \sqrt{e^{*}_{c}}$, which for a 
degenerate Fermi gas reduces to $(T_c - T) / T_c$.\ \ The excitation energy normalized to the mass of the fragmenting remnant, 
$e^*$ in MeV/nucleon, excludes collective effects \cite{hauger_prc}.\ \ The location of the critical point, $e^{*}_{c}$, and values 
of the critical exponents, $\sigma$ and $\tau$, were determined previously \cite{gilkes_gamma}, \cite{elliott_sigma}, 
\cite{hauger_prl}, \cite{hauger_prc}, \cite{elliott_scaling}.\ \ Fitting $n^{scaled}_{A}$ as a function of ${\epsilon}^{scaled}$ for 
$\epsilon \ge 0$ and leaving $c_0$ and $\exp \left[ A{\Delta}{\mu} / \sqrt{e^*} \right]$ as free parameters in 
Eq.~(\ref{fisher_droplet}) gives $c_0 = 6.4 \pm 0.6$ MeV (via $E^* = a T^2$ with $a = A_0 / 13$) and 
$\exp \left[ A {\Delta}{\mu} / \sqrt{e^*} \right] = 0.8 \pm 0.1$, {\it i.e.} the bulk term is consistent with 
${\Delta}{\mu} \approx 0$.\ \ The surface energy coefficient $c_0$ is of a somewhat different nature than the semiempirical mass formula 
parameter ($a_s \sim 17$ MeV for $T = 0$, $\rho = {\rho}_0$) or estimates for low density nuclear systems 
($a_s \sim 6$ MeV for $T \sim 3$ MeV, $\rho \sim {\rho}_0 / 3$) \cite{hirsch}.\ \ The coefficient $c_0$ is temperature independent; 
the temperature dependence is given as $c_0 \epsilon$.

Since the FDM has been shown to contain the features of reducibility and thermal scaling and since the scaling inherent in the 
FDM describes percolation, it should be possible to observe reducibility and thermal scaling in percolation cluster distributions.

\begin{figure} [ht]
\centerline{\psfig{file=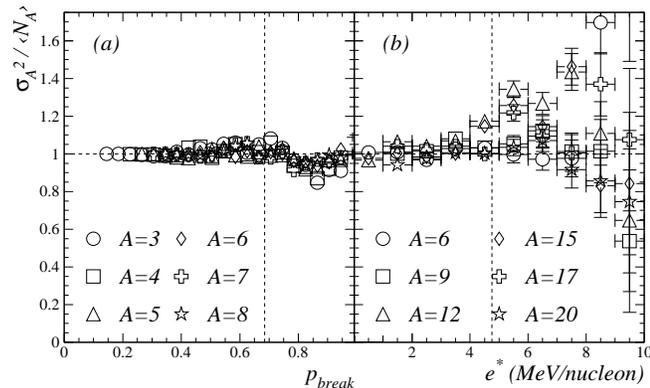,width=8.5cm,angle=0}}
\caption{Ratio of the variance to the mean number of clusters (fragments) of size (mass) $A$ as a function of $p_{break}$ 
($e^*$) for percolation (a) (Au multifragmentation (b)).\ \ Error bars are statistical.\ \ Vertical dashed lines show 
the location of the critical point.}
\label{perc_ratio}
\end{figure}

To address the question of reducibility in percolation, cluster multiplicity distributions for bins in $p_{break}$ are considered.\ \ 
The ratio of the variance to the mean, ${\sigma}^{2}_{A} / \left< N_A \right>$, of the multiplicity distribution for each cluster 
of size $A$ is an indicator of the nature of the distribution.\ \ In Fig.~\ref{perc_ratio}, such a ratio is shown as a function 
of $p_{break}$.\ \ The observed ratio is near one (Poissonian limit) over the range of $p_{break}$.\ \ Within experimental 
errors, similar behavior is observed for the Au multifragmentation data.

\begin{figure} [ht]
\centerline{\psfig{file=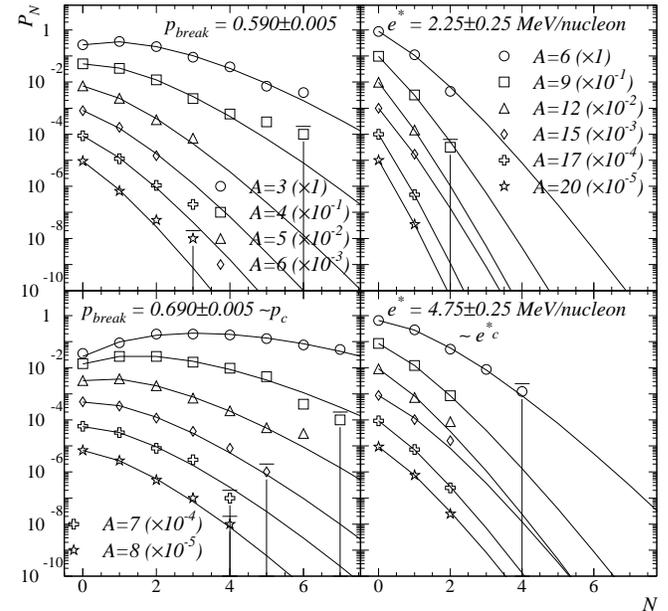,width=8.5cm,angle=0}}
\caption{Multiplicity distributions, $P_N$, for clusters (fragments) of size (mass) $A$ as a function of $N$ for bins in 
$p_{break}$ ($e^*$) for percolation (Au multifragmentation) on the left (right).\ \ The lines are Poissonian distributions 
calculated from the measured $\left< N_A \right>$.}
\label{perc_pois}
\end{figure}

Examples of multiplicity distributions with Poissonian curves calculated from percolation $\left< N_A \right>$ are shown in 
the left panel of Fig.~\ref{perc_pois}.\ \ Poissonian distributions reproduce percolation cluster distributions over two or 
three orders of magnitude for all $A$ values; Poissonian reducibility is present in percolation.

Fig.~\ref{perc_pois} also shows the multiplicity distributions for Au multifragmentation compared to the calculated Poissonian 
curves.\ \ The agreement between the measured and computed distributions confirms the presence of reducibility in the Au 
multifragmentation data.

\begin{figure} [ht]
\centerline{\psfig{file=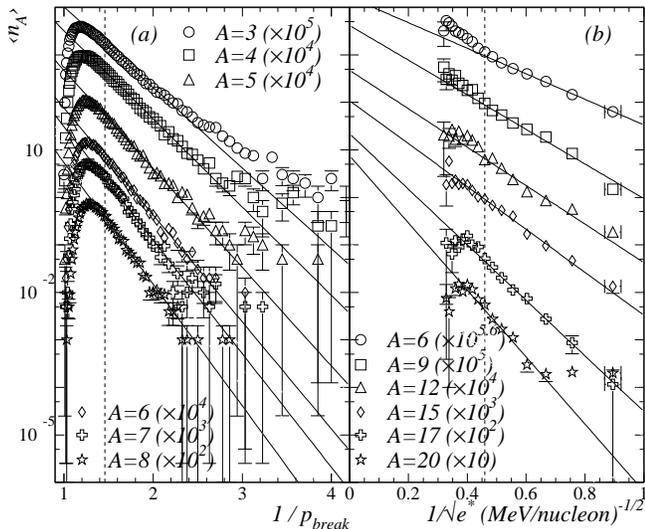,width=8.5cm,angle=0}}
\caption{Normalized average cluster (fragment) multiplicity as a function of $1/p_{break}$ ($1 / \sqrt{e^*}$) for clusters 
(fragments) of different size (mass), $A$, for percolation (a) (Au multifragmentation (b)).\ \ Solid lines show Arrhenius 
fits.\ \ Vertical dashed lines show the location of the critical point.}
\label{perc_arrh}
\end{figure}

To verify thermal scaling in percolation, the average yield of clusters of size $A$ and its dependence on $p_{break}$ are 
considered.\ \ The presence of thermal scaling should manifest itself through a Boltzmann factor.\ \ Again, the substitution 
$p_{break}$ for $T$ is made in accordance with standard percolation theory and $\ln \left< n_A \right>$ is plotted as a function of 
$1 / p_{break}$ (Arrhenius plot).\ \ See Fig.~\ref{perc_arrh}.\ \ In most cases the Arrhenius plots for individual clusters of size 
$A$ are linear over two orders of magnitude.\ \ Thus, thermal scaling is verified for percolation.\ \ The observations that 
reducibility and thermal scaling are already present in such a simple model suggest that they are deeply rooted fundamental features 
of multifragmentation processes rather than being epiphenomena of complex systems.

\begin{figure} [ht]
\centerline{\psfig{file=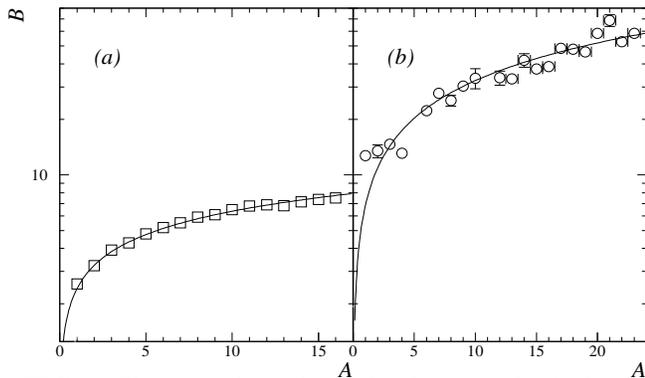,width=8.5cm,angle=0}}
\caption{The power law relationship between the Arrhenius barrier, $B$, and the cluster size (fragment mass) $A$ for percolation (a) 
(multifragmentation (b)).\ \ The solid line is a fit to the points with $A > 1$.}
\label{perc_barr}
\end{figure}

The Boltzmann factor indicates that the slope of an Arrhenius plot represents the barrier $B$ associated with the production of 
a cluster.\ \ For an interpretation of $B$ in percolation, the Boltzmann factor is equated with Eq.~(\ref{fisher_droplet}) yielding 
a power law relating $B$ to the size of a cluster: $B = c_0 A^{\sigma}$ when ${\Delta}{\mu} \approx 0$.\ \ Fitting the extracted 
barriers $B$ (slopes in Fig.~\ref{perc_arrh}) as a function of $A$ gives an exponent equal to $0.42 \pm 0.02$ in agreement with the 
accepted value of ${\sigma} = 0.45$ for $3$D percolation.\ \ See Fig.~\ref{perc_barr}.\ \ The constant of proportionality of the 
power law gives another measure of the surface energy coefficient $c_0 = 2.42 \pm 0.03$.

When the Arrhenius analysis is performed on the Au multifragmentation data, the results are qualitatively similar, but 
quantitatively distinct.\ \ Arrhenius fits of $\ln \left< n_A \right>$, now plotted against $1 / \sqrt{e^*}$, are linear over 
an order of magnitude or more.\ \ See Fig.~\ref{perc_arrh}.\ \ The barriers extracted here can be converted into units of MeV 
via the energy-temperature relation for a degenerate Fermi gas.\ \ Following the same analysis as for percolation, a fit was made 
of $B$ vs. $A$ (see Fig.~\ref{perc_barr}) which yielded $\sigma = 0.68 \pm 0.03$, in agreement with previously determined EOS Au 
multifragmentation values \cite{elliott_sigma}, \cite{elliott_scaling}, \cite{elliott_stat} and $c_0 = 6.8 \pm 0.5$ MeV.

In summary, the above effort illustrates:
\begin{itemize}
 \item the presence of reducibility and thermal scaling in the FDM, percolation and the EOS Au multifragmentation data (the latter 
       two shown empirically);
 \item the relationship between the FDM and percolation via a scaling analysis that also yields an estimate of the surface energy 
       coefficient;
 \item that the barriers obtained from percolation follow a power-law dependence on cluster size with an exponent that agrees with 
       the accepted $3$D percolation value and gives another (consistent) estimate of the surface energy coefficient;
 \item the collapse of the EOS Au fragment distributions in accordance with the FDM yielding an estimate of the surface energy 
       coefficient of a low density nuclear system;
 \item that the barriers obtained from EOS Au multifragmentation data follow a power-law dependence on fragment mass with an exponent 
       near the expected value ($2/3$) and close to the 3D Ising universality class value ($0.64$) and gives another (consistent) 
       estimate of the surface energy coefficient.
\end{itemize}

This work was supported in part by the U.S. Department of Energy and by the National Science Foundation.


\begin{thebibliography}{}

\bibitem{fisher}
	M. E. Fisher, Physics {\bf 3}, 255 (1967).

\bibitem{stauffer_aharony}
	D. Stauffer and A. Aharony, ``Introduction to Percolation Theory'', 2nd ed. (Taylor and Francis, London, 1992).

\bibitem{finn}
	J. E. Finn {\it et al}., Phys. Rev. Lett. {\bf 49}, 1321 (1982).

\bibitem{campi_2}
	X. Campi, Phys. Lett. B {\bf 208}, 351 (1988).

\bibitem{bauer_1}
	W. Bauer {\it et al}., Phys. Lett. B 150, {\bf 53} (1985).

\bibitem{gilkes_gamma}
	M. L. Gilkes, {\it et al}., Phys. Rev. Lett. {\bf 73}, 1590 (1994).

\bibitem{elliott_sigma}
	J. B. Elliott {\it et al}., Phys. Lett. B, {\bf 381}, 24 (1996).

\bibitem{hauger_prl}
	J. A. Hauger {\it et al}., Phys. Rev. Lett. {\bf 77}, 235 (1996).

\bibitem{hauger_prc}
	J. A. Hauger {\it et al}., Phys. Rev. C {\bf 57}, 764 (1998).

\bibitem{elliott_scaling}
	J. B. Elliott {\it et al}., Phys. Lett. B. {\bf 418}, 35 (1998).

\bibitem{moretto_rev}
	L. G. Moretto {\it et al}., Phys. Rep. {\bf 287}, 249 (1997).

\bibitem{beaulieu}
	L. Beaulieu {\it et al}., Phys. Rev. Lett. {\bf 81}, 770 (1998).

\bibitem{toke}
	J. Toke {\it et al}., Phys. Rev. C {\bf 56}, R1683 (1997).

\bibitem{phair}
	L. G. Moretto {\it et al}., Phys. Rev. C {\bf 60}, 031601 (1999).

\bibitem{frenkel}
	J. Frenkel, ``Kinetic Theory of Liquids'', Oxford University Press (1946).

\bibitem{mayer}
	J. E. Mayer and M. G. Mayer, ``Statistical Mechanics'', John Wiley, New York (1940).

\bibitem{nakanishi}
	H. Nakanishi and H. E. Stanley, Phys. Rev. B {\bf 22}, 2466 (1980).

\bibitem{stanley}
	H. E. Stanely {\it et al}., Physica A {\bf 266}, 5 (1999).

\bibitem{elliott_perc_2}
	J. B. Elliott {\it et al}., Phys. Rev. C {\bf 55}, 1319, (1997).

\bibitem{hirsch}
	A. S. Hirsch {\it et al}., Phys. Rev. C {\bf 29}, 508 (1984).

\bibitem{elliott_stat}
	J. B. Elliott {\it et al}., in progress.

\end{thebibliography}
\end{document}